\documentclass[twocolumn,showpacs,preprintnumbers,amsmath,amssymb]{revtex4}
\usepackage{amsfonts}
\usepackage{mathrsfs}
\usepackage{amsmath}
\usepackage{graphicx}
\usepackage{dcolumn}
\usepackage{bm}
\usepackage{hyperref}

\begin{document} 
\title{Quantum Coherence Effects in Four-level Diamond Atomic
System} \author{Bao-Quan Ou}
  \email{bqou@nudt.edu.cn}
\author{Lin-Mei Liang}
 \email{nmliang@nudt.edu.cn}
\author{Cheng-Zu Li}
\affiliation{Department of Physics, Science College, National
University of Defense Technology,  Changsha, 410073, China .}
\date{\today}
\begin{abstract}

 A symmetric four-level closed-loop $\diamondsuit$ type (the diamond structure) atomic
system driven by four coherent optical fields is investigated. The
system shows rich quantum interference and coherence features. When
symmetry of the system is broken, interesting phenomena such as
single and double dark resonances appear. As a result, the double
electromagnetically induced transparency effect is generated, which
will facilitate the implementation of quantum phase gate operation.
 \end{abstract}

\pacs{42.50.Gy,42.50.Nn,42.50.Ex} 

\maketitle

\section{\label{sec:Introd}Introduction}

Driven by coherent optical fields, atomic system demonstrates
abundant interference and coherent effects. Interesting effects such
as the coherent population trapping state(CPT)\cite{QuantOpt},
electromagnetically induced transparency (EIT)\cite{harris,
RMP000633} and laser without inversion (LWI)\cite{PRL2813,
OptCom499}, have been theoretically and experimentally studied by
many research groups in the world\cite{nature594, nature731,
nature837, PRL2053, PRL173601}. By investigating these coherent
effects, the nature of quantum interference and coherence is
comprehended further. For example, in the three-level $\Lambda$
atomic system, the transition from
\textquotedblleft{dark}\textquotedblright state to the excited state
is canceled via the quantum interference of two-photon resonance
transition, makes the optical medium
\textquotedblleft{transparent}\textquotedblright, thus the EIT
effect is generated\cite{QuantOpt, RMP000633}. This EIT effect has
many attractive applications in quantum optics, such as the
multi-wave mixing\cite{OL769, OL804, PRL126303}, enhancement of
nonlinear susceptibilities\cite{PRAzhu, PRA041801}. More
interestingly, the EIT effect has been found applications in quantum
information science, such as the photon information storage and
release in an atomic assemble\cite{PRL783}, correlated photon pairs
generation\cite{nature731} and even the entanglement of remote
atomic assembles\cite{nature828}, which form the building blocks of
the quantum communication and the quantum computation.

An ideal level structure atomic system with appreciate interference
and coherence features will bring great help to implement quantum
states operation, and facilitate the information transferring
between different systems. The most popular three-level systems in
quantum optics are the $\Lambda$, $V$ and the $\Xi$
systems\cite{RMP000633,nature594, nature731}, these atomic systems
have the CPT, EIT and LWI effects under different actions of driving
fields. More complicated levels structure of atomic system will
generate more interesting effects. For example, doubly
electromagnetically induced transparency (double EIT) in the double
$\Lambda$ four-level system\cite{OL769, OL804}. Other multi-level
atomic systems, such as the tripod type four-level
system\cite{PRA032317}, the inverted-Y system\cite{PRA062319} and
the N-type system\cite{OL1936, JMO1559}, have been explored and
found interesting application in the demonstration of quantum logic
gate. Looking for an \emph{ideal} atomic system which is rich in
quantum interference and coherence effects and easy to demonstrate
experimentally is still under way.

\begin{figure}
\scalebox{0.95}{\includegraphics[width=\columnwidth]{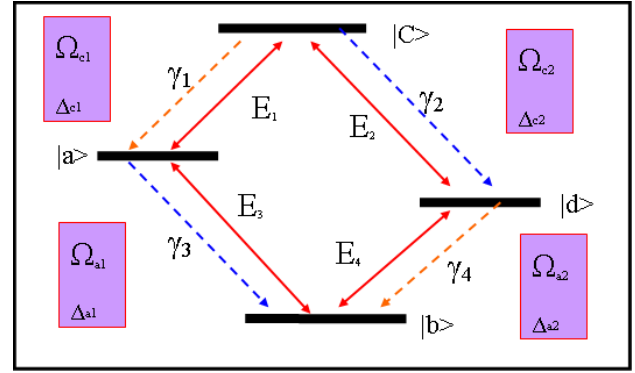}}
\caption{\label{fig:levelscheme} Scheme of a four-level diamond
atomic system. }
 \end{figure}

Here we are investigating the quantum interference and coherence
characteristics of a highly symmetric four-level atomic system, the
$\diamondsuit$ (or the diamond structure) system\cite{PRA3225,
PRA053409, PRL203001} driven by four coherent optical fields. This
kind of atomic structure has been proposed to study the interaction
of laser phase and the steady state\cite{PRA053409}, and to generate
ultra-violet laser gain through the LWI effect\cite{PRA3225,
OptCom499}. While our investigation on the system is to seek the
application to the implementation of quantum phase gate, an
attractive operation based on cross-phase modulation effect or
equally the double EIT effect(or double-dark
resonance)\cite{PRA062319}. Inspired by the job of M.D. Lukin
\textit{et al.}\cite{PRA3225, OptCom499}, we generalize their atomic
model to a symmetric and closed structure, and discuss several more
possible ways of obtaining the multi-dark resonance effect. Under
different ways of configuration of the coherent driving fields, the
double-dark states are generated, thus the double EIT effect is
attained. Due to the symmetric structure of the atomic system, all
the coherent driving fields can be assigned to be the probe, trig
and couple fields alternately, and the EIT effect on the probe and
trig fields simultaneously occur. The transparency of the probe and
the trig fields are both controlled by the couple field, they will
vary from transparent to opaque with the strength of the couple
field changing. This flexibility of arrangement the coherent driving
fields will facilitate the implementation of quantum phase gate.

 This paper is organized as follow: firstly in
 section\ref{sec:TransitAnalys}, we use the dressed state method to
 study the quantum interference effects of the system driven by four
 optical fields, and reveal the mechanics of generating the dark
 states; then in section\ref{sec:MasterEq} we solve the master
 equations and investigate varying laws of
 the susceptibility of the probe and the trig fields, brings up the appreciated
 double EIT effects. Finally in section\ref{sec:Diss}, we discuss the
 potential application to quantum phase gate of the atomic system
 and summarize our results.

\section{\label{sec:TransitAnalys}The model: Transition Analysis of
Four-level Diamond System}
 The four-level system we are focusing on
is shown in Fig.\ref{fig:levelscheme}, where the four atomic levels
form a so-called diamond ($\diamondsuit$) or a closed-loop
configuration\cite{PRA053409,PRL203001}. The upper state is
$|c\rangle$ state, the bottom state is $|b\rangle$ state, and the
two intermediate states are $|a\rangle$ and $|d\rangle$. As shown in
the figure, the four-level system are driven by four coherent
optical fields $E_{1-4}$. These four coherent fields are coupling
the four dipole-allowed transitions of the atomic system, whose Rabi
frequencies are $\Omega_{c1}, \Omega_{c2}, \Omega_{a1}$ and
$\Omega_{a2}$, respectively. The decay processes are denoted by
decay rates $\gamma_i$ ($i=1,2,3,4$) in different decay channels.
Within the rotating wave approximation (RWA) and in the interaction
picture, the process of coherently driving the diamond atomic system
can be described by the following Hamiltonian
matrix\cite{QuantOpt}\cite{QuantOpt2}:
 \begin{eqnarray} H=-\hbar\
\begin{bmatrix} \Delta_{a1} & \Omega_{a1} & \Omega^{*}_{c1} & 0\\
\Omega^{*}_{a1} & 0 & 0 & \Omega^{*}_{a2} \\ \Omega_{c1} & 0 &
\Delta_{a1}+\Delta_{c1} & \Omega^{*}_{c2}\\ 0 & \Omega_{a2} &
\Omega^{*}_{c2} & \Delta_{a2}
 \end{bmatrix}
 \label{eq:Hcoh},
\end{eqnarray}
  where the sequence of state vectors is $(|a\rangle,
|b\rangle, |c\rangle, |d\rangle)^T$. For simplicity, the Rabi
frequencies $\Omega_{ci}$ and $\Omega_{ai}$ ($i=1,2$) are chosen to
be real in the following, i.e., $\Omega_{ci}=\Omega^*_{ci}$ and
$\Omega_{ai}=\Omega^*_{ai}$. The detunings $\Delta_{ai}$ and
$\Delta_{ci}$ ($i=1,2$) are defined as
$\Delta_{c1}=\omega_{c1}-\omega_{ca}$,
$\Delta_{c2}=\omega_{c2}-\omega_{cd}$,
$\Delta_{a1}=\omega_{a1}-\omega_{ab}$,
$\Delta_{a2}=\omega_{a2}-\omega_{db}$, and
$\omega_{ij}=\omega_{i}-\omega_{j}$, where $\omega_{i}, (i=a,b,c,d)$
denote the frequency of atomic level state,  $\omega_{ci}$ and
$\omega_{ai}$ ($i=1,2$) are the frequencies of the corresponding
coherent fields. With the Hamiltonian matrix Eq.(\ref{eq:Hcoh}), we
can analyze the quantum interference and quantum coherence behaviors
of the system under the dressed state picture. Firstly, we choose
the driving field $E_2$ which couples the $|c\rangle \Leftrightarrow
|d\rangle$ transition to be the probe field, and the left three
coherent fields to be coupling and driving fields. For simplicity,
we only consider the situation of zero detunings $\Delta_{k}=0$($
k=a1,a2,c1,c2$), that is, all the coupling and driving fields are
tuned on resonance with the corresponding transitions, respectively.
The four dressed states are:
\begin{widetext} \footnotesize \begin{eqnarray} \label{eq:DressSt1}
|D1\rangle&=&\Omega_{a1}[\frac{\Omega_{a1}(y-\sqrt{z})}{(w-\sqrt{z})\sqrt{z+x\sqrt{z}}}|a\rangle
+\frac{\sqrt{2}}{2}\sqrt{\frac{y-\sqrt{z}}{z+x\sqrt{z}}}|b\rangle
-\frac{\Omega_{c1}}{\sqrt{z-w\sqrt{z}}}|c\rangle
+\frac{\Omega_{a2}}{\sqrt{z+x\sqrt{z}}}]|d\rangle\\
|D2\rangle&=&\Omega_{a1}[\frac{\Omega_{a1}(y-\sqrt{z})}{(w-\sqrt{z})\sqrt{z+x\sqrt{z}}}|a\rangle
-\frac{\sqrt{2}}{2}\sqrt{\frac{y-\sqrt{z}}{z+x\sqrt{z}}}|b\rangle
+\frac{\Omega_{c1}}{\sqrt{z-w\sqrt{z}}}|c\rangle
+\frac{\Omega_{a2}}{\sqrt{z+x\sqrt{z}}}]|d\rangle\\
|D3\rangle&=&\frac{\Omega_{a1}(y+\sqrt{z})}{\sqrt{2(w+y)z+2(wx+2wy-xy)\sqrt{z}}}|a\rangle
+\frac{w+\sqrt{z}}{2\sqrt{w\sqrt{z}+z}}|b\rangle
+\frac{\Omega_{c1}\Omega_{a1}}{\sqrt{w\sqrt{z}+z}}|c\rangle
+\frac{\Omega_{a2}(w+\sqrt{z})}{\sqrt{2(w+y)z+2(wx+2wy-xy)\sqrt{z}}}|d\rangle\\
|D4\rangle&=&\frac{\Omega_{a1}(y+\sqrt{z})}{\sqrt{2(w+y)z+2(wx+2wy-xy)\sqrt{z}}}|a\rangle
-\frac{w+\sqrt{z}}{2\sqrt{w\sqrt{z}+z}}|b\rangle
-\frac{\Omega_{c1}\Omega_{a1}}{\sqrt{w\sqrt{z}+z}}|c\rangle
+\frac{\Omega_{a2}(w+\sqrt{z})}{\sqrt{2(w+y)z+2(wx+2wy-xy)\sqrt{z}}}|d\rangle
 \label{eq:DressSt4},
\end{eqnarray} \end{widetext}
where $w=\Omega^2_{a2}-\Omega^2_{c1}+\Omega^2_{a1}$,
$x=\Omega^2_{a2}-\Omega^2_{c1}-\Omega^2_{a1}$,
$y=\Omega^2_{a2}+\Omega^2_{c1}+\Omega^2_{a1}$, and
$z=y^2-4\Omega^2_{a2}\Omega^2_{c1}$. The corresponding eigen-energy
are: $\epsilon_1=-\sqrt{\frac{y-\sqrt{z}}{2}}$
,$\epsilon_2=\sqrt{\frac{y-\sqrt{z}}{2}}$
,$\epsilon_3=-\sqrt{\frac{y+\sqrt{z}}{2}}$
,$\epsilon_4=\sqrt{\frac{y+\sqrt{z}}{2}}$, respectively. By
analyzing these dressed states, we are willing to see that the
diamond atomic system contains plenty of quantum interference and
quantum coherence effects. Intuitively, there are one photon
excitation process, two-photon resonance excitation, and even
three-photon resonance excitation process in this atomic system
driven by four coherent fields, that is, many possible quantum
transition paths exist in the closed-loop system. These different
quantum transition paths will interfere with each other, and result
in the interesting effects such as electromagnetically induced
transparency, the laser without inversion and the enhancement of
nonlinear refractive index and etc. As is known to all, the dark
state is the heart of the above quantum interference and coherence
effects. While the dressed states
Eq.(\ref{eq:DressSt1}-\ref{eq:DressSt4}) will help us to find the
desired dark states. Obviously all four dressed states
$|D1\rangle$-$|D4\rangle$ are containing all the four bare states
elements $|a\rangle, |b\rangle, |c\rangle, |d\rangle$, indicating
the absence of the so-called
\textquotedblleft{dark}\textquotedblright  state. In order to
acquire the dark state which is decoupled from the high level state
$|c\rangle$ of the probing transition, we can adjust the three
coherent driving fields. Considering that the atomic structure is a
closed-loop type, a highly symmetric structure, the probe field can
connect to all the four bare states elements through all possible
exciting paths, the connection between the high level state
$|c\rangle$ and other dressed states will remain. If the symmetry is
broken, some dressed states may disconnect from the probe
transition, then the dark states will appear. In the following we
will break the symmetry of the system by removing one driving field,
and inspect the resulted dark states and their interesting results.

There are three coherent optical fields acting on the diamond system
except the probe field, we can remove any one of them to reach the
aim of symmetry breaking, that is, the closed-loop system will
become an open-loop one, and three branches of the $\diamondsuit$
are left. For convenience, we denote \emph{case 1} of removing the
driving field $E_3$, \emph{case 2} of removing the driving field
$E_4$, and \emph{case 3} of removing the driving field $E_1$.
Firstly, we take \emph{case 1} into account, where the $E_3$ field
is removed (equally take $\Omega_{a1}\rightarrow 0$), then the
dressed states Eq.(\ref{eq:DressSt1}-\ref{eq:DressSt4}) turn into
the following form:
 \begin{eqnarray} \label{eq:SimpDressSt1}
|\mathcal {D}_{1}\rangle&=& \frac{1}{\sqrt{\Omega^2_{a2}+\Omega^2_{c1}}} (\Omega_{c1}|b\rangle+\Omega_{c1}|d\rangle),\nonumber \\
|\mathcal {D}_{2}\rangle&=& \frac{1}{\sqrt{\Omega^2_{a2}+\Omega^2_{c1}}} (-\Omega_{c1}|b\rangle+\Omega_{c1}|d\rangle),\nonumber \\
|\mathcal {D}_3\rangle&=&\frac{1}{\sqrt{2}}(|b\rangle+|d\rangle), \nonumber \\
|\mathcal {D}_4\rangle&=&\frac{1}{\sqrt{2}}(-|b\rangle+|d\rangle),
\end{eqnarray}
where the strength of the left driving fields $E_1, E_4$ satisfy the
condition: $\Omega_{c1}\neq \Omega_{a2}$; while for situation
$\Omega_{c1} = \Omega_{a2}$, the dressed states
Eq.(\ref{eq:DressSt1}-\ref{eq:DressSt4}) become degenerated:

\begin{eqnarray} \label{eq:SimpDressSt12}
|\mathcal {D}_{1,3}\rangle&=&\frac{1}{\sqrt{2}}(|b\rangle+|d\rangle), \nonumber \\
|\mathcal{D}_{2,4}\rangle&=&\frac{1}{\sqrt{2}}(-|b\rangle+|d\rangle),
\end{eqnarray}

\begin{figure}
 \begin{center}
  \includegraphics[bb=0 0 412 197, width=0.6\columnwidth]{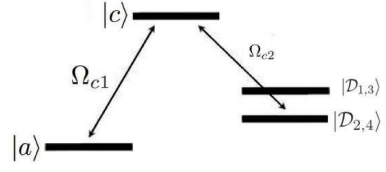}
 \end{center}
 \caption{\label{fig:dress1}Transition picture of
diamond atomic system under the coherent pumping of three fields
$E_1, E_2, E_4$}.
\end{figure}

The above results Eq.(\ref{eq:SimpDressSt1}) and
Eq.(\ref{eq:SimpDressSt12}) show an interesting feature: the four
dressed states $|\mathcal {D}_{1,2,3,4}\rangle$ do not have the
upper level state element $|c\rangle$ state of the probing
transition. Thus the two states $|\mathcal {D}_{2,4}\rangle$ form
the desired dark states. Now the probing field exciting the
$|c\rangle\Leftrightarrow|d\rangle$ transition is coupling with the
two dark states $|\mathcal {D}_{2,4}\rangle$ and form the double
dark resonances, quantum interference will occur for these two
transition paths. More interestingly, when the strength of the two
driving fields are equal: $\Omega_{c1}=\Omega_{a2}$, the four dress
states $|\mathcal {D}_{1,2,3,4}\rangle$ will degenerate into two
states as shown in Eq.(\ref{eq:SimpDressSt12}), thus the two dark
states of Eq.(\ref{eq:SimpDressSt1}) merge into one dark state, and
the interaction of dark states will disappear. This particular
feature will bring up interesting phenomenon, which will be
discussed in the following. While all the four dressed states
contain the bare state $|b\rangle$ component, thus the probe field
excites a two-photon resonance excitation: $|c\rangle
\Leftrightarrow |b\rangle$. The transition picture is shown in
Fig.\ref{fig:dress1}.

\begin{figure}
 \begin{center}
  \includegraphics[width=0.6\columnwidth]{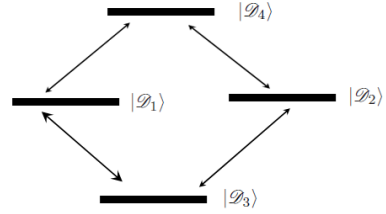}
 \end{center}
 \caption{\label{fig:dress2}Transition picture of
diamond atomic system under the coherent pumping of three fields
$E_1, E_2, E_3$}.
\end{figure}

For finding other dark states resonance, we now discuss the second
way of symmetry broken of the diamond structure atomic system, the
\emph{case 2}. Instead of removing the driving field $E_3$, we now
withdraw the field $E_4$, equally take $\Omega_{a2}\rightarrow 0$.
As a result, the dressed states
Eq.(\ref{eq:DressSt1}-\ref{eq:DressSt4}) become:

 \begin{eqnarray}
\label{eq:SimpDressSt2}
|\mathscr{D}_1\rangle&=&\frac{1}{\sqrt{2(\Omega^2_{a1}+\Omega^2_{c1})}}(\Omega_{c1}|b\rangle-\Omega_{a1}|c\rangle)+\frac{1}{\sqrt{2}}|d\rangle,\nonumber \\
|\mathscr{D}_2\rangle&=&\frac{1}{\sqrt{2(\Omega^2_{a1}+\Omega^2_{c1})}}(-\Omega_{c1}|b\rangle+\Omega_{a1}|c\rangle)+\frac{1}{\sqrt{2}}|d\rangle,\nonumber \\
|\mathscr{D}_3\rangle&=&\frac{1}{\sqrt{2}}|a\rangle+\frac{1}{\sqrt{2(\Omega^2_{a1}+\Omega^2_{c1})}}(\Omega_{a1}|b\rangle+\Omega_{c1}|c\rangle),\nonumber \\
|\mathscr{D}_4\rangle&=&\frac{1}{\sqrt{2}}|a\rangle-\frac{1}{\sqrt{2(\Omega^2_{a1}+\Omega^2_{c1})}}(\Omega_{a1}|b\rangle+\Omega_{c1}|c\rangle).
\end{eqnarray}

 From the form of the above dressed states $|\mathscr{D}_i\rangle$
$(i=1,2,3,4)$, it's clear that they are all coupled with the state
$|c\rangle$, thus the dark state does not show up at the moment. If
we take the two coherent driving field $E_1$ and $E_3$ equally
interacting on the atomic system, that is,
$\Omega_{c1}=\Omega_{a1}$, then the dressed state
Eq.(\ref{eq:SimpDressSt2}) become a much simple form:
 $|\mathscr{D}_1\rangle= \frac{1}{2}(|b\rangle-|c\rangle)+\frac{1}{\sqrt{2}}|d\rangle$,
  $|\mathscr{D}_2\rangle= \frac{1}{2}(-|b\rangle+|c\rangle)+\frac{1}{\sqrt{2}}|d\rangle$,
  $|\mathscr{D}_3\rangle=\frac{1}{\sqrt{2}}|a\rangle+\frac{1}{2}(|b\rangle+|c\rangle)$,
and
$|\mathscr{D}_4\rangle=\frac{1}{\sqrt{2}}|a\rangle-\frac{1}{2}(|b\rangle+|c\rangle)$.
The four dressed states are containing the transition
$|c\rangle\Leftrightarrow |b\rangle$, corresponding the two-photon
resonances transition, and states $|\mathscr{D}_{1,2}\rangle$
include transition $|d\rangle\Leftrightarrow |b\rangle$,
corresponding a three-photon resonance transition. These transitions
interact with each other, results in abundant quantum interference
effects, the dressed states transition picture is shown in
Fig.\ref{fig:dress2}. This kind of situation is quite the same as
that discussed in Ref.\cite{PRA3225}. If we further discuss the
situation in the limit of vanishing perturbation of the coherent
driving fields $\Omega_{a1}\rightarrow 0$ (or
$\Omega_{c1}\rightarrow 0$), the dressed states
Eq.(\ref{eq:SimpDressSt2}) will turn out to be the single dark
resonance (or double dark resonances). Thus the quantum interference
effects induced by dark resonance (or interacting dark resonances)
will appear.

Likewise, in the symmetry broken \emph{case 3}, the coherent field
$E_1$ is removed, then the diamond atomic system is driven by the
field $E_{3,4}$ and the probe field $E_2$. As a result, the dressed
states Eq.(\ref{eq:DressSt1}-\ref{eq:DressSt4}) now become:

 \begin{eqnarray}
\label{eq:SimpDressSt3}
 |\mathfrak{D}_{1}\rangle&=&\frac{1}{\sqrt{2(\Omega^2_{a1}+\Omega^2_{a2})}}(-\Omega_{a2}|a\rangle+\Omega_{a1}|d\rangle)-\frac{1}{\sqrt{2}}|c\rangle,\nonumber \\
 |\mathfrak{D}_{2}\rangle&=&\frac{1}{\sqrt{2(\Omega^2_{a1}+\Omega^2_{a2})}}(-\Omega_{a2}|a\rangle+\Omega_{a1}|d\rangle)+\frac{1}{\sqrt{2}}|c\rangle,\nonumber \\
 |\mathfrak{D}_{3}\rangle&=&\frac{1}{\sqrt{2(\Omega^2_{a1}+\Omega^2_{a2})}}(\Omega_{a1}|a\rangle+\Omega_{a2}|d\rangle)+\frac{1}{\sqrt{2}}|b\rangle,\nonumber \\
 |\mathfrak{D}_{4}\rangle&=&\frac{1}{\sqrt{2(\Omega^2_{a1}+\Omega^2_{a2})}}(\Omega_{a1}|a\rangle+\Omega_{a2}|d\rangle)-\frac{1}{\sqrt{2}}|b\rangle,
\end{eqnarray}

\begin{figure}
\begin{center} \includegraphics[width=0.6\columnwidth]{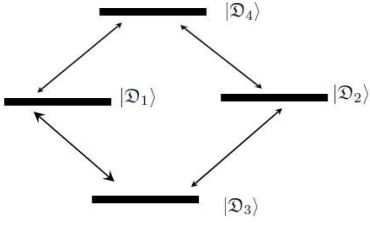}
\end{center} \caption{\label{fig:dress3}Transition picture of
diamond atomic system under the coherent pumping of three fields
$E_2, E_3, E_4$}.
\end{figure}

Obviously, the dressed states $|\mathfrak{D}_{3,4}\rangle$  are
decoupled from the excited state $|c\rangle$, and form the desired
dark states. As is discussed in the symmetry broken \emph{case 2},
here the two-photon resonances transition $|a\rangle \Leftrightarrow
|d\rangle$ and the three-photon resonance transition $|c\rangle
\Leftrightarrow |a\rangle$ are involved in the dressed states. The
dressed states transition picture is drawn as Fig.\ref{fig:dress3}.

The above dressed states analysis has shown the variety quantum
interference and coherent effects in the $\diamondsuit$ system.
However, it is just a kind of qualitative study of the system, for
we don't take the detuning of each coherent fields into account,
neither the decay rates, and these two factors are also important
that they will modify the coherent characters of the system. In
order to learn more subtle behaviors of the atomic system, further
investigations on the variation law of level-population terms, probe
susceptibility terms and multi-photon resonance excitation terms are
needed. In the next section, we will study the master equations and
the corresponding solutions to the system, and reveal the
interesting quantum interference and coherence effects.

\section{\label{sec:MasterEq} Master Equations and
Solutions of the $\diamondsuit$ system }
 We denote the density
matrix for the atomic system by $\rho$, the master equation for the
density matrix $\rho$ is\cite{QuantOpt}\cite{QuantOpt2}:
 \begin{equation}
\frac{\partial}{\partial t}\rho=-\frac{i}{\hbar}[H,\rho]+\mathcal
{L}\rho
 \label{eq:Master},
\end{equation}
where the first term on the right side is the coherent driving part,
the corresponding Hamiltonian $H$ takes the form of
Eq.(\ref{eq:Hcoh}). The second term is the relaxation part, i.e. the
spontaneous decay process. In the present four-level $\diamondsuit$
atomic system, there are four decay channels, from upper levels
decay to lower levels, as shown in Fig.\ref{fig:levelscheme}. The
corresponding relaxation operator describing the decay process is:

 \begin{eqnarray}
 \mathcal{L}\rho&=&\mathcal {L}_{ca}\rho+\mathcal {L}_{cd}\rho+\mathcal
{L}_{ab}\rho+\mathcal {L}_{db}\rho\nonumber\\
&=&\frac{\gamma_1}{2}(2|a\rangle
\langle{c}|\rho|c\rangle\langle{a}|-|c\rangle\langle{c}|\rho-\rho|c\rangle\langle{c}|)\nonumber\\
&& +\frac{\gamma_2}{2}(2|d\rangle
\langle{c}|\rho|c\rangle\langle{d}|-|c\rangle\langle{c}|\rho-\rho|c\rangle\langle{c}|)\nonumber\\
&& +\frac{\gamma_3}{2}(2|b\rangle
\langle{a}|\rho|a\rangle\langle{b}|-|a\rangle\langle{a}|\rho-\rho|a\rangle\langle{a}|)\nonumber\\
&& +\frac{\gamma_4}{2}(2|b\rangle
\langle{d}|\rho|d\rangle\langle{b}|-|d\rangle\langle{d}|\rho-\rho|d\rangle\langle{d}|)
 \label{eq:Hdecay}.
\end{eqnarray}

 With the Hamiltonian Eq.(\ref{eq:Hcoh}) and Eq.(\ref{eq:Hdecay}), we
have the motional equations for the atomic density matrix elements:
{\footnotesize
\begin{eqnarray}
 \label{eq:rhoaa}
\dot{\rho}_{aa}&=&-\gamma_3\rho_{aa}+\gamma_1\rho_{cc}+i\Omega_{a1}(\rho_{ba}-\rho_{ab})+i\Omega_{c1}(\rho_{ca}-\rho_{ac}) \nonumber \\
\dot{\rho}_{bb}&=&\gamma_3\rho_{aa}+\gamma_4\rho_{dd}+i\Omega_{a1}(\rho_{ab}-\rho_{ba})+i\Omega_{a2}(\rho_{db}-\rho_{bd}) \nonumber \\
\dot{\rho}_{cc}&=&-(\gamma_1+\gamma_2)\rho_{cc}+i\Omega_{c1}(\rho_{ac}-\rho_{ca})+i\Omega_{c2}(\rho_{dc}-\rho_{cd}) \nonumber \\
\dot{\rho}_{dd}&=&\gamma_2\rho_{cc}-\gamma_4\rho_{dd}+i\Omega_{a2}(\rho_{bd}-\rho_{db})+i\Omega_{c2}(\rho_{cd}-\rho_{dc}) \\
\label{eq:enclose}
\rho_{aa}&+&\rho_{bb}+\rho_{cc}=1 \\
\label{eq:rhoab}
\dot{\rho}_{ab}&=&(i\Delta_{a1}-\frac{\gamma_3}{2})\rho_{ab}+i\Omega_{a1}(\rho_{bb}-\rho_{aa})-i\Omega_{a2}\rho_{ad}+i\Omega_{c1}\rho_{cb}
\nonumber
 \\&&\\
\label{eq:rhoac}
\dot{\rho}_{ac}&=&-[i\Delta_{c1}+\frac{1}{2}(\gamma_1+\gamma_2+\gamma_3)]\rho_{ac}+i\Omega_{c1}(\rho_{cc}-\rho_{aa}) \nonumber \\
&&-i\Omega_{c2}\rho_{ad}+i\Omega_{a1}\rho_{bc}\\
\label{eq:rhoad}
\dot{\rho}_{ad}&=&[i(\Delta_{a1}-\Delta_{a2})-\frac{1}{2}(\gamma_3+\gamma_4)]\rho_{ad}-i\Omega_{a2}\rho_{ab}-i\Omega_{c2}\rho_{ac} \nonumber \\
&&+i\Omega_{a1}\rho_{bd}+i\Omega_{c1}\rho_{cd} \\
\label{eq:rhobc}
\dot{\rho}_{bc}&=&-[i(\Delta_{c1}+\Delta_{a1})+\frac{1}{2}(\gamma_1+\gamma_2)]\rho_{bc}-i\Omega_{c1}\rho_{ba}+i\Omega_{a1}\rho_{ac} \nonumber \\
&&+i\Omega_{a2}\rho_{dc}-i\Omega_{c2}\rho_{bd}\\
\label{eq:rhobd}
\dot{\rho}_{bd}&=&-[i\Delta_{a2}+\frac{\gamma_4}{2}]\rho_{bd}-i\Omega_{a2}(\rho_{bb}-\rho_{dd})+i\Omega_{a1}\rho_{ad}-i\Omega_{c2}\rho_{bc}\nonumber
 \\&&\\
\label{eq:rhocd}
\dot{\rho}_{cd}&=&[i(\Delta_{a1}+\Delta_{c1}-\Delta_{a2})-\frac{1}{2}(\gamma_1+\gamma_2+\gamma_4)]\rho_{cd}
\nonumber \\
&&-i\Omega_{c2}(\rho_{cc}-\rho_{dd})+i\Omega_{c1}\rho_{ad}-i\Omega_{a2}\rho_{cb}
\end{eqnarray}}
where Eq.(\ref{eq:enclose}) expresses the
conservation of probability for the closed diamond system.

\begin{figure}
 \begin{center}
\includegraphics[width=0.8\columnwidth]{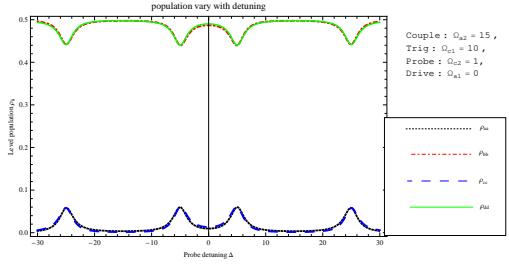}
 \end{center}
\caption{\label{fig:popu1}Bare states population
$\rho_{ii}(i=a,b,c,d)$ vary with probe field detuning
$\Delta_{c2}=\Delta$. The dotted line is $\rho_{aa}$ and the
dash-dotted line is $\rho_{bb}$, the dash line is $\rho_{cc}$, and
the solid line is $\rho_{dd}$. The parameters are: $\Omega_{a2}=15,
\Omega_{c2}=1, \Omega_{c1}=10, \Omega_{a1}=0$, all the decay rates
take $\gamma_i=1(i=1,2,3,4)$, all detunings take zero value except
the probing one.}
 \end{figure}

\begin{figure}
 \begin{center}
\includegraphics[width=0.8\columnwidth]{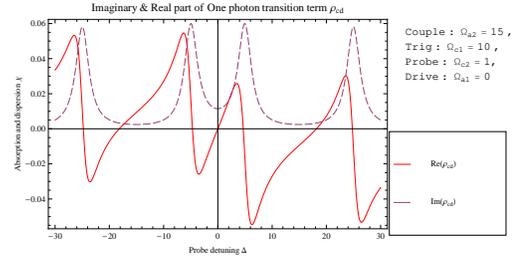}
 \end{center}
\caption{\label{fig:probe1} Probe susceptibility term $\rho_{cd}$
varies with field detuning $\Delta$. The parameters are the same as
those in Fig.\ref{fig:popu1}. The solid line is the real part of
$\rho_{cd}$, which describes the dispersion of the probe field; the
dashed is the imaginary part of $\rho_{cd}$, which describes the
absorption of the probe field $E_2$.}
 \end{figure}

\begin{figure}
 \begin{center}
 \includegraphics[width=0.85\columnwidth]{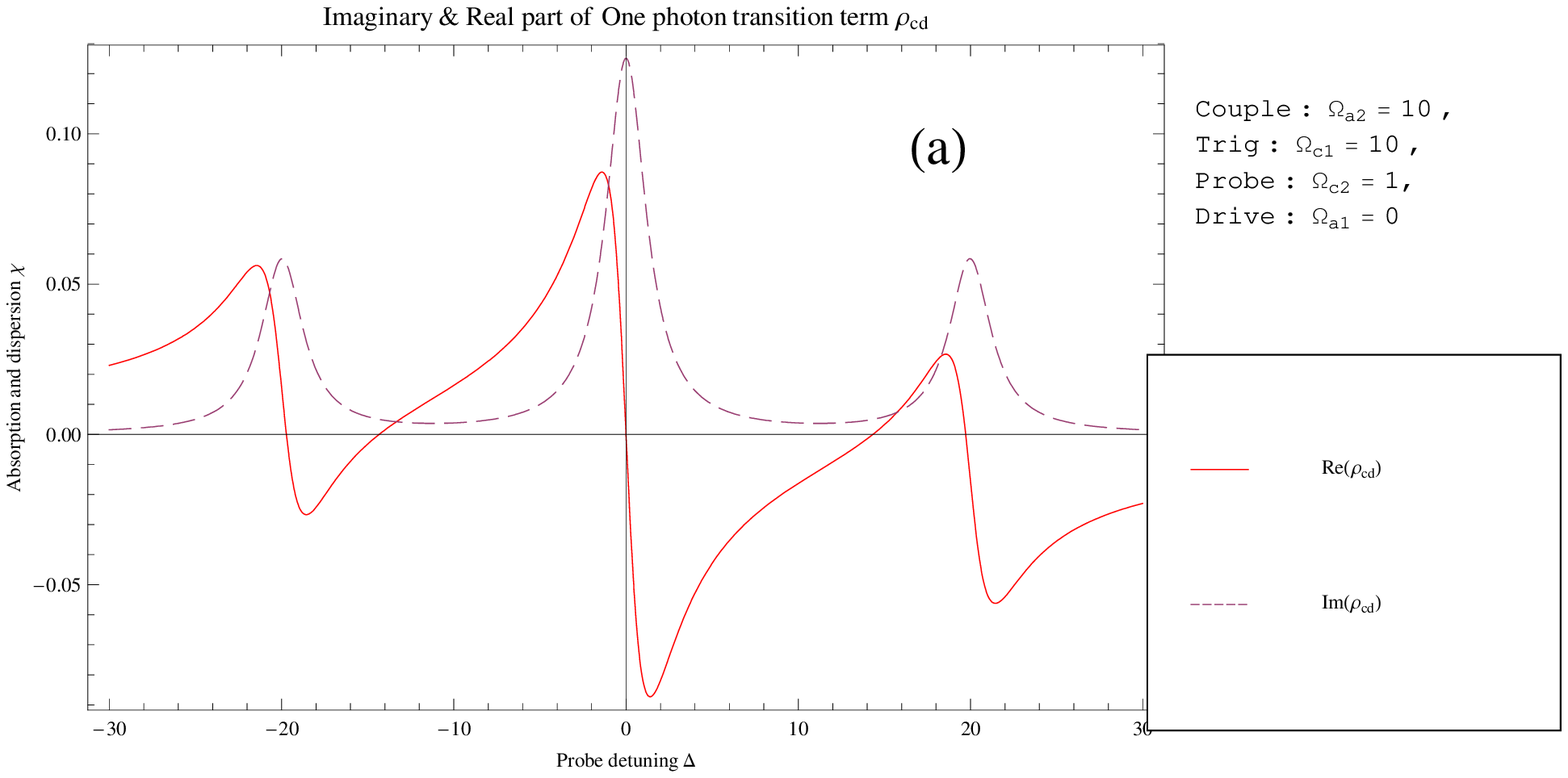}
\includegraphics[width=0.85\columnwidth]{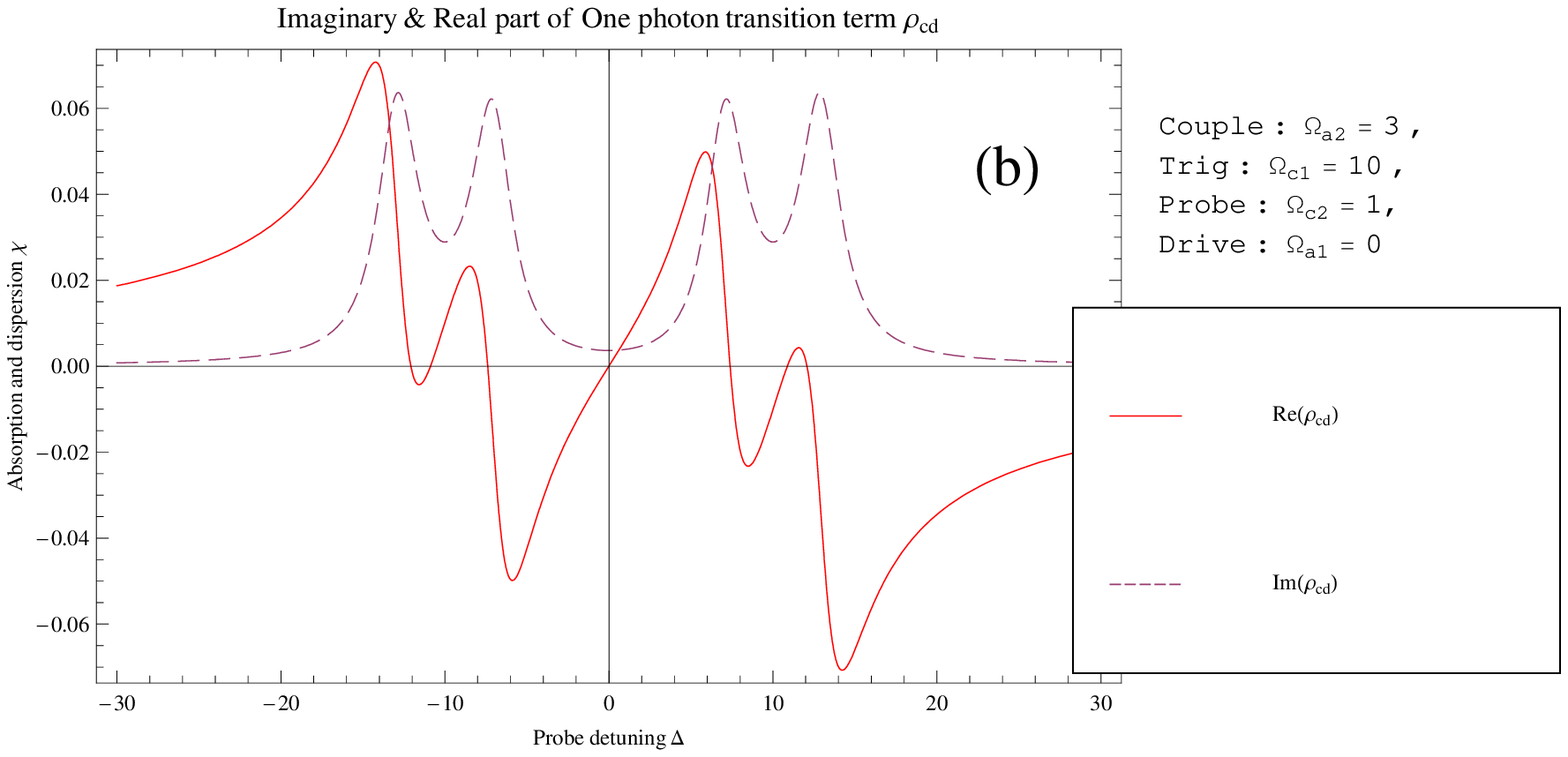}
 \end{center}
\caption{\label{fig:probe10} Probe susceptibility term $\rho_{cd}$
varies with field detuning $\Delta$. With the driving field takes
$\Omega_{a2}=10$ (figure a), or $\Omega_{a2}=3$ (figure b). The
other parameters are the same as those in Fig.\ref{fig:popu1}. }
 \end{figure}

\begin{figure}
 \begin{center}
\includegraphics[width=0.8\columnwidth]{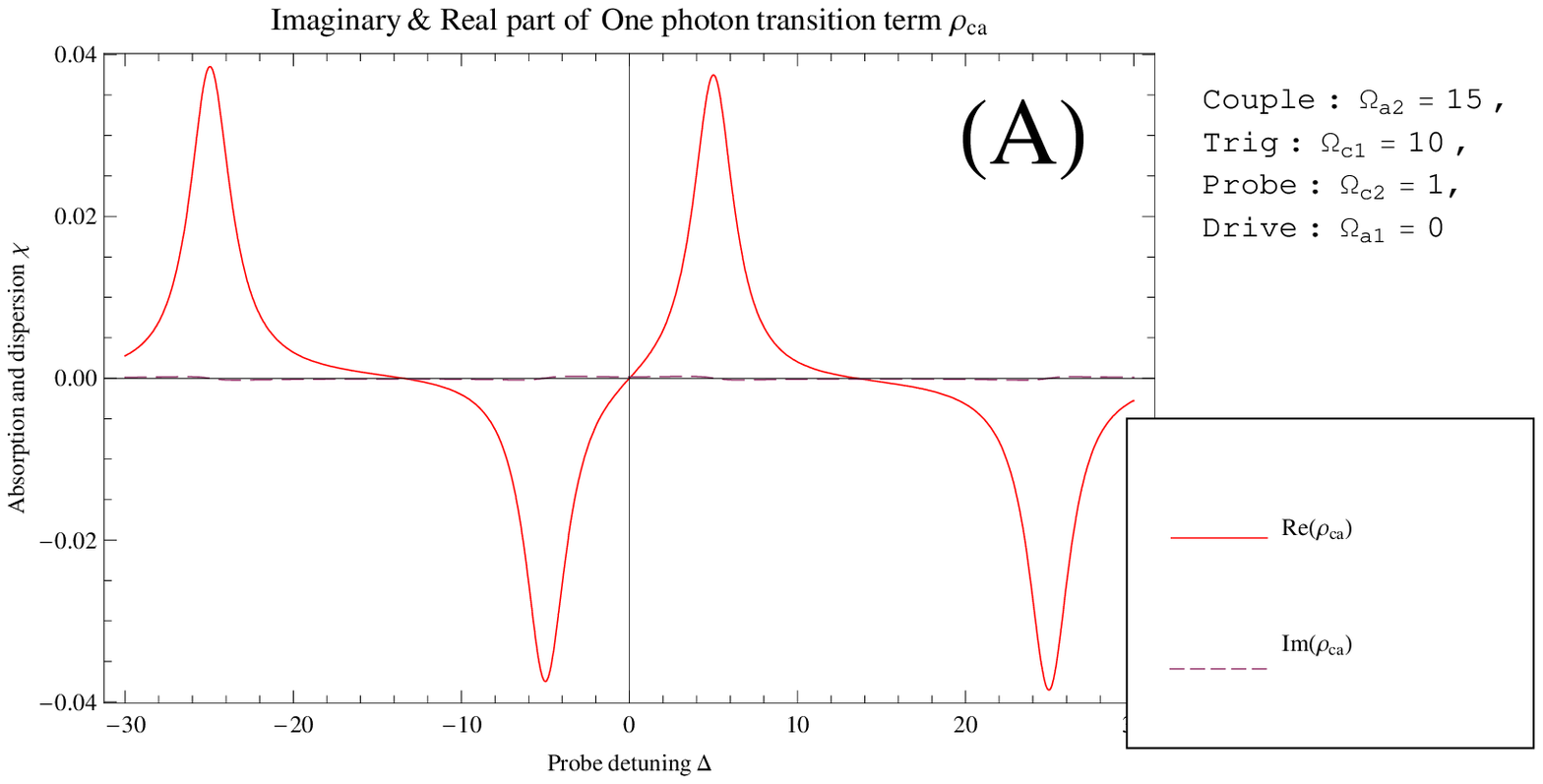}
\includegraphics[width=0.8\columnwidth]{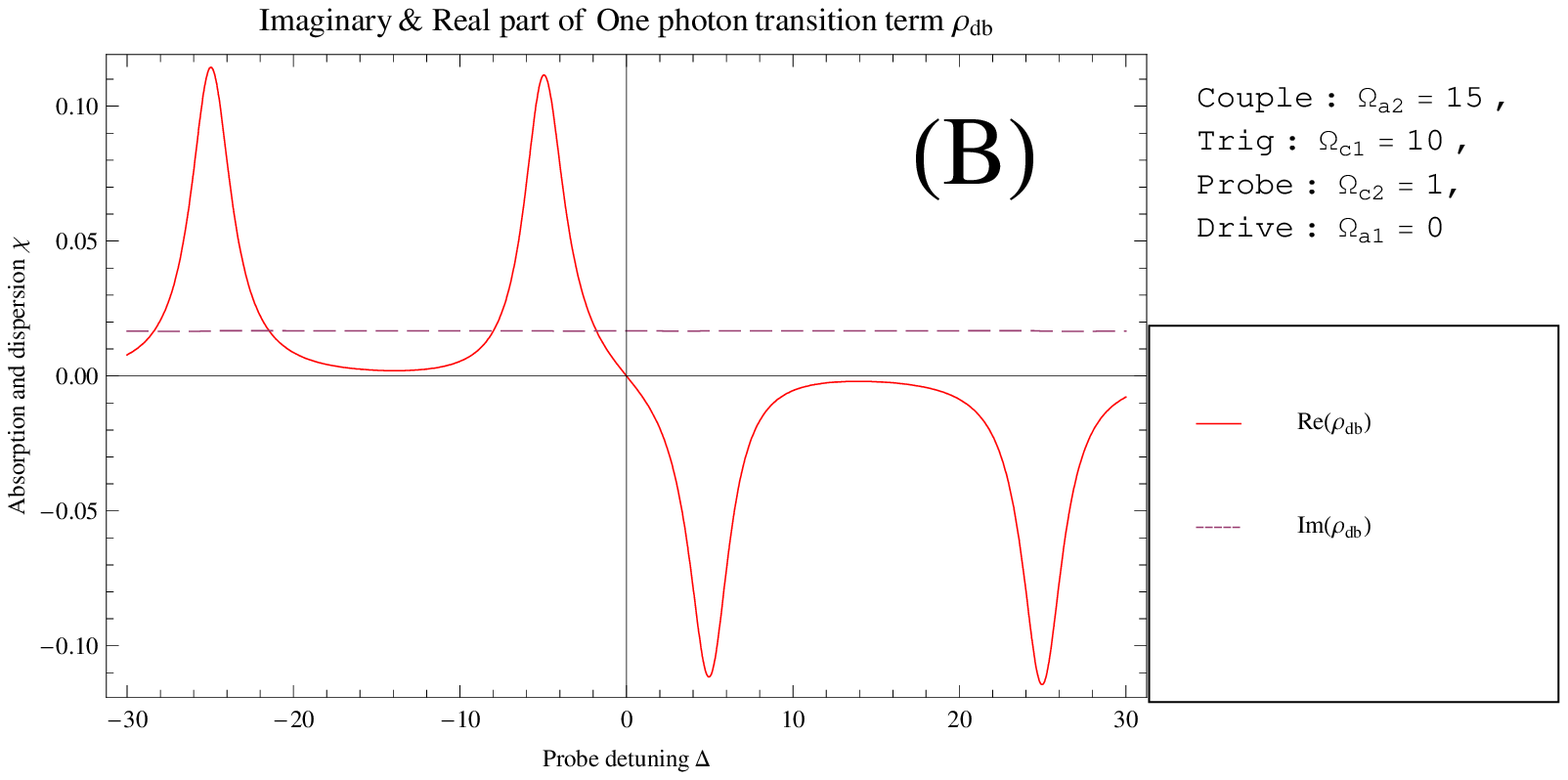}
 \end{center}
\caption{\label{fig:trig1} Trig(figure A) and couple(figure B)
transition terms $\rho_{ca}$ and $\rho_{db}$ vary with field
detuning $\Delta$. The parameters are the same as those in
Fig.\ref{fig:popu1}. }
 \end{figure}

\begin{figure}
  \begin{center}
\includegraphics[width=0.8\columnwidth]{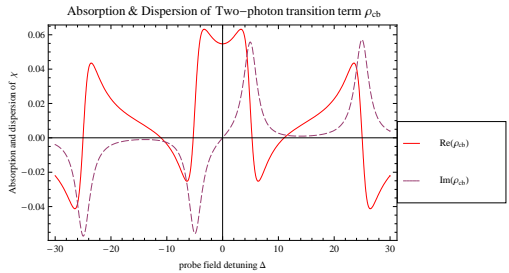}
  \end{center}
\caption{\label{fig:2photonRes1} Two-photon resonance excitation
term $\rho_{cb}$ varies with probe field detuning $\Delta$. The
parameters are the same as those in Fig.\ref{fig:popu1}. The solid
line is the real part of $\rho_{cb}$; the dash is the imaginary part
of $\rho_{cb}$.}
 \end{figure}

 According to the dressed states analysis, we also discuss three
kinds of symmetry broken situations here, that is, the closed-loop
structure of the system will become an open one, by removing one of
the three driving optical fields. Firstly we discuss \emph{case 1},
where the field $E_3$ is removed, that is, $\Omega_{a1}=0$. For
convenience, the left three coherent fields are denoted as the probe
field $E_2$, the trig(signal) field $E_1$ and the couple field
$E_4$, respectively. The solutions to the master equations
Eq.(\ref{eq:rhoaa}-\ref{eq:rhocd}) are shown in
Fig.(\ref{fig:popu1}-\ref{fig:2photonRes1}).

From the level population Fig.\ref{fig:popu1}, it's clear that
 the four bare level-states are divided into two groups,
 group 1 comprises state $|a\rangle$ and state $|c\rangle$, group 2
 comprises state $|b\rangle$ and state $|d\rangle$.
 State population in the same group are nearly equal, which matches
 well with the dressed states analysis result Eq.(\ref{eq:SimpDressSt1}).
 Varying with the probe field detuning $\Delta_{c2}$, all four
 state-population come to peaks or troughs, revealing the situation
 of population transferring between the four bare states, as well as the
 absorption of the coherent driving field and the probe fields.

In order to learn more details of optical fields interacting with
the atomic system, we investigate the probe field's absorption and
dispersion features vary with the detuning $\Delta$, the result is
shown in the Fig.\ref{fig:probe1}. Comparing the Fig.\ref{fig:popu1}
and the Fig.\ref{fig:probe1}, it's easy to find that the level
population and the absorption of the probe field come to the same
peaks and troughs at the same values of independent variable
detuning $\Delta$, which demonstrates that the variation of the
absorption of the probe field results in the perturbation of
level-population.

And the dispersion and absorption features of the probe field
deserve special attention. From Fig.\ref{fig:probe1} it's clear that
the normal and anomalous dispersion alternatively appear varying
with the detuning $\Delta$, considering the behavior of the
absorption part together, the probe field goes through a
three-windows electromagnetically induced transparency (EIT)
process. For investigating the EIT effect further, we decrease the
strength of the couple field and study the resulted variation of the
probe field. When the strength of the couple field decreases, the
depth and width of the middle transparency window become smaller
than those in Fig.\ref{fig:probe1}. When the strength of the couple
field equals to that of the trig field, that is,
$\Omega_{a2}=\Omega_{c1}=10$, the probe field lose the transparency
character and turns out to be a absorption peak, and the anomalous
dispersion curve turns to a normal one, as shown in
Fig.\ref{fig:probe10}(a). However, if we continue to decrease the
strength of the couple field, the middle transparency window appears
once again, and the two sideward transparency windows become less
transparency, as is shown in Fig.\ref{fig:probe10}(b), where the
couple field is much weaker than the trig field: $\Omega_{a2}=3$.
From the dressed states analysis
Eq.(\ref{eq:SimpDressSt1},\ref{eq:SimpDressSt12}), we know that the
two dark states $|\mathcal{D}_2\rangle$ and $|\mathcal{D}_4\rangle$
will become degenerate for condition $\Omega_{c1}=\Omega_{a2}$, thus
the on resonance EIT feature of the probe field will vanish.

Another interesting feature of the dark resonances of the system
comes from the absorption and dispersion behaviors of the trig field
$E_1$, which is represented by the imaginary and real part of the
$\rho_{ca}$ transition term, as is shown in Fig.\ref{fig:trig1}(A).
It's obvious that the trig field becomes transparency on resonance,
thus the trig field also has the EIT effect. While the couple field
$E_4$ goes through a steady absorption process, as is shown in
Fig.\ref{fig:trig1}(B). If the strength of couple field is equal to
that of trig field $E_1$, the EIT effect for the trig field will
also disappear. Combining the above double-EIT behaviors of the
probe field $E_2$ and the trig field $E_1$, this system will be
useful on realizing the quantum phase
gate\cite{PRA032317,PRA062319}, which is very important in quantum
information science.

The quantum interference will be revealed more clearly if we
investigate the two-photon resonance excitation term $\rho_{cb}$,
which is plotted in Fig.\ref{fig:2photonRes1}. Obviously, the
absorption behaviors of the two-photon resonance transition are
identical to those of the probe transition, except that in the
negative detuning region, the absorption character of the two-photon
excitation become negative ones, which means gain characteristic.
And if we adjust the strength of couple field $E_4$ equal to that of
trig field $E_1$, the two absorption peaks on both sides of zero
detuning will merge together and cancel each other, indicates that
the two-photon resonance excitation disappear on resonance, which
also confirm the above dressed states analysis.

\begin{figure}
 \begin{tabular}{|c|c|}\hline
a \includegraphics[width=0.45\columnwidth]{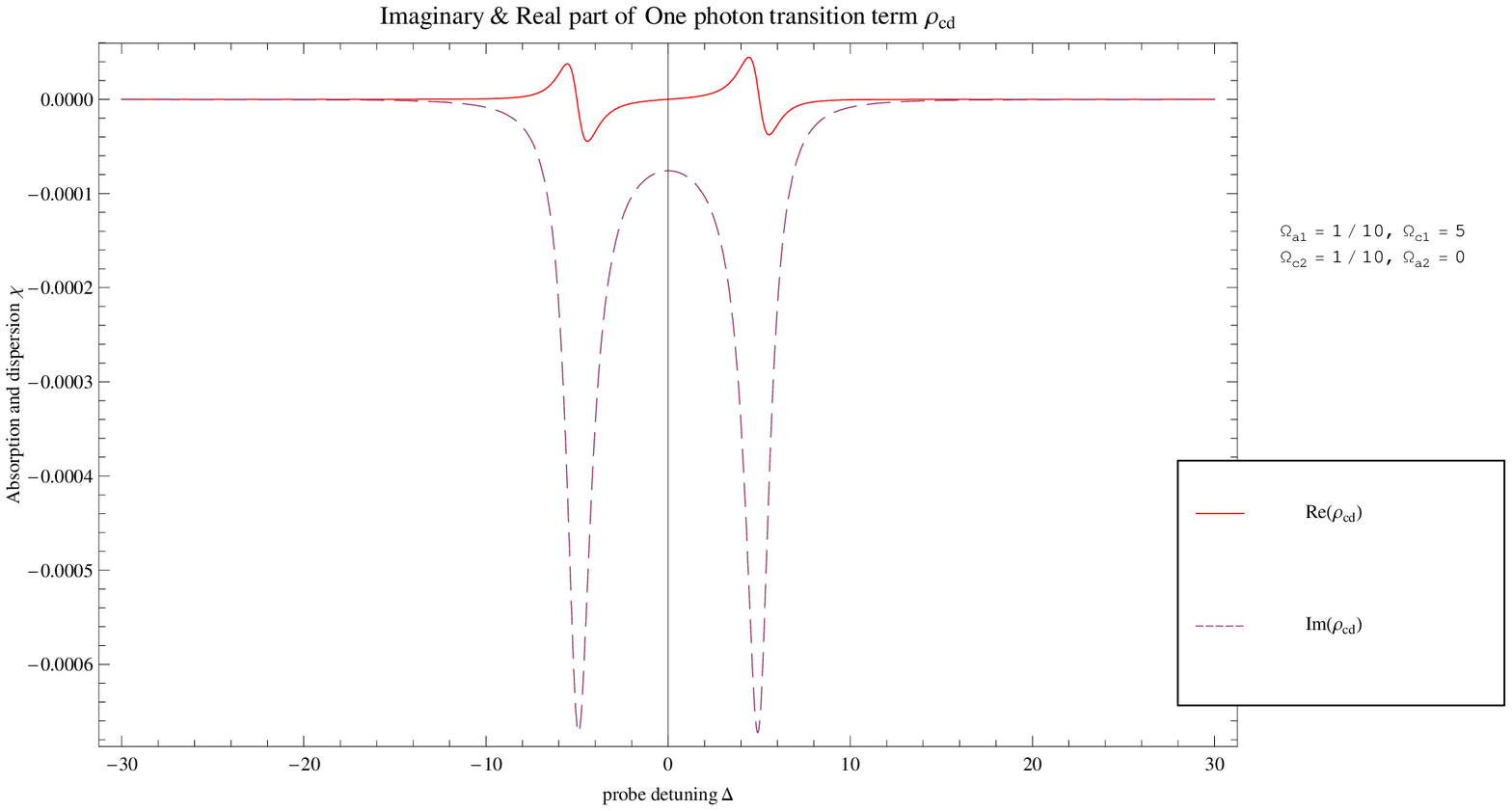}&
 a'\includegraphics[width=0.45\columnwidth]{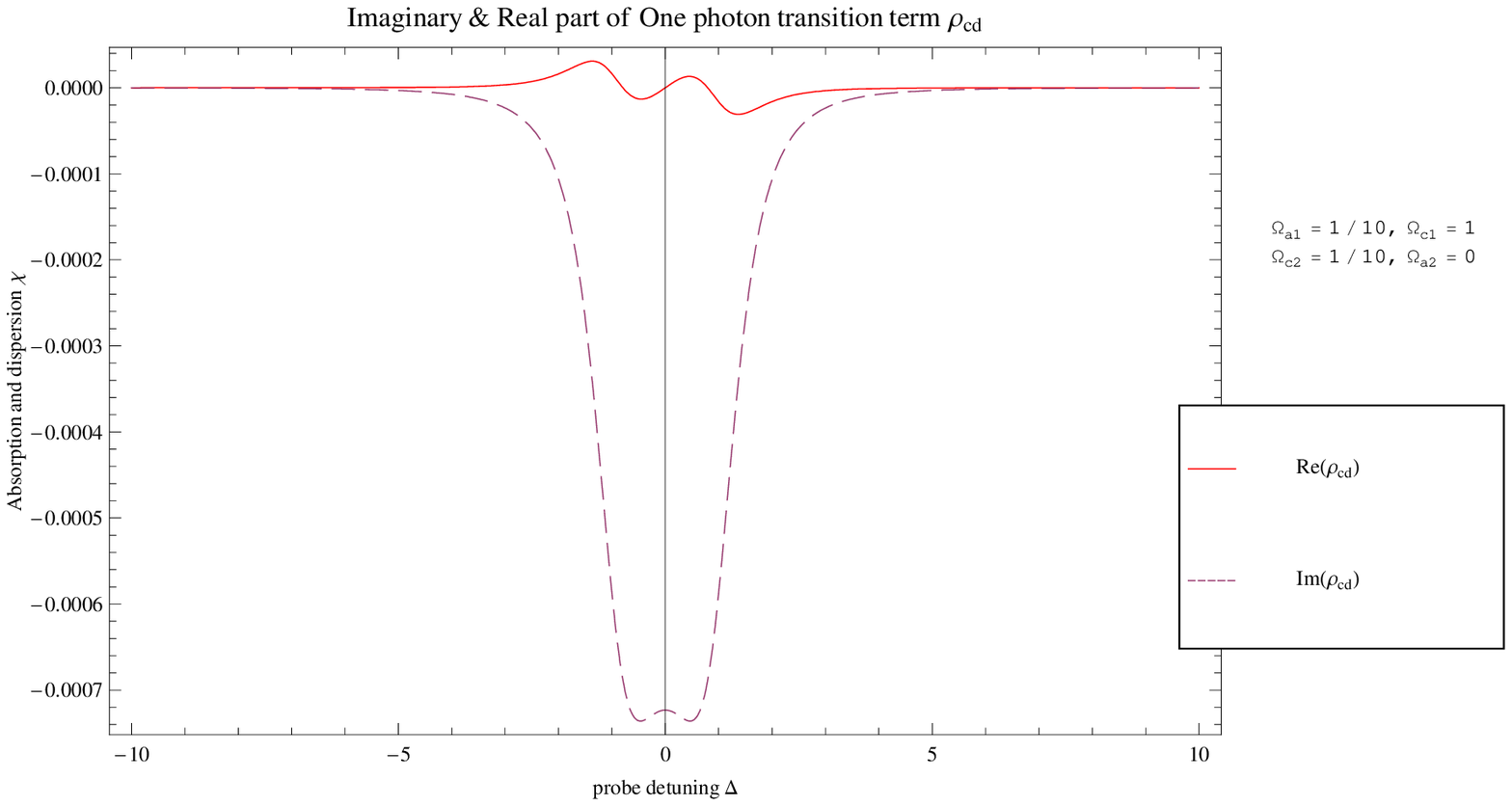}\\ \hline
  b\includegraphics[width=0.45\columnwidth]{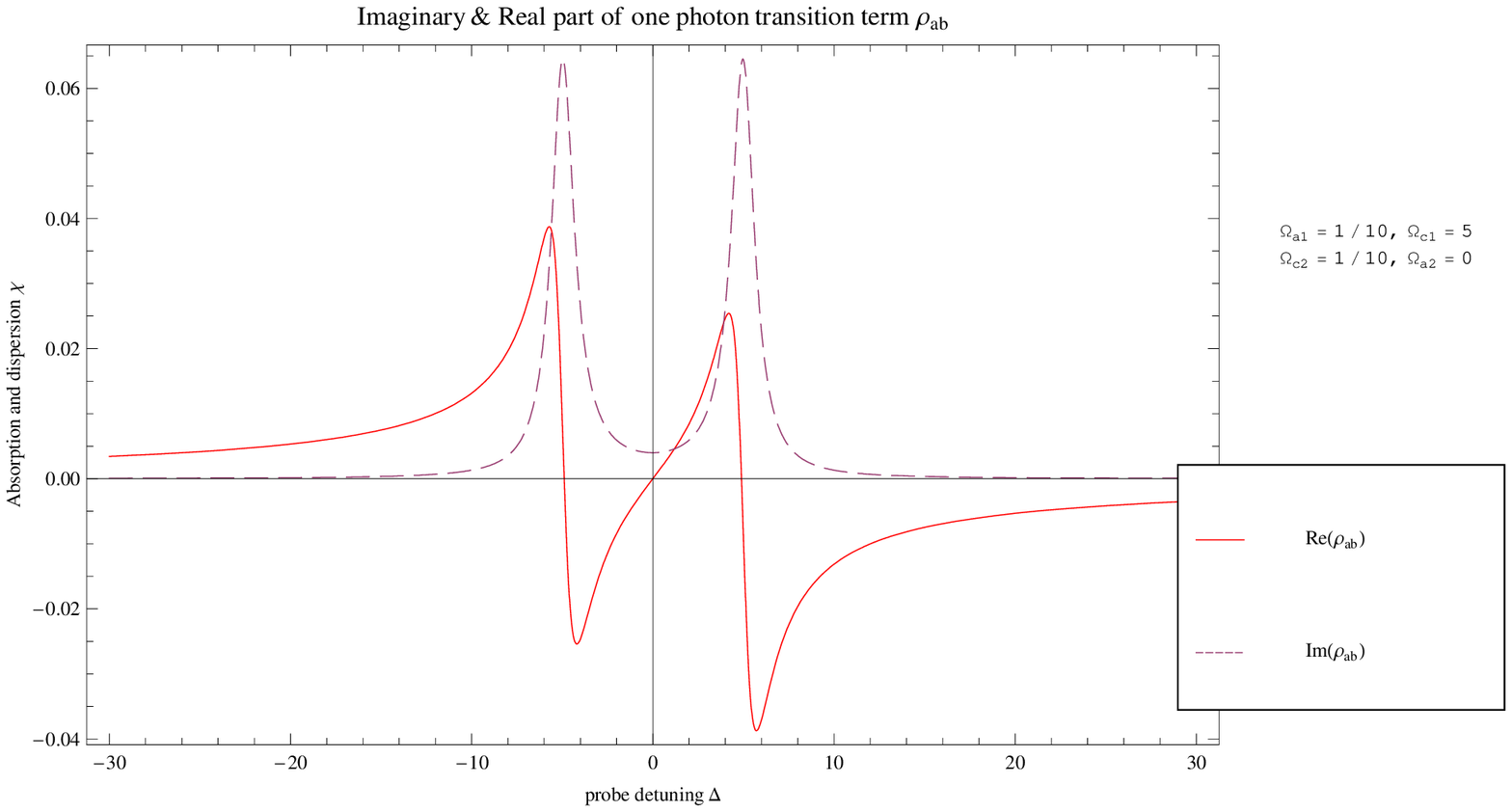}&
 b'\includegraphics[width=0.45\columnwidth]{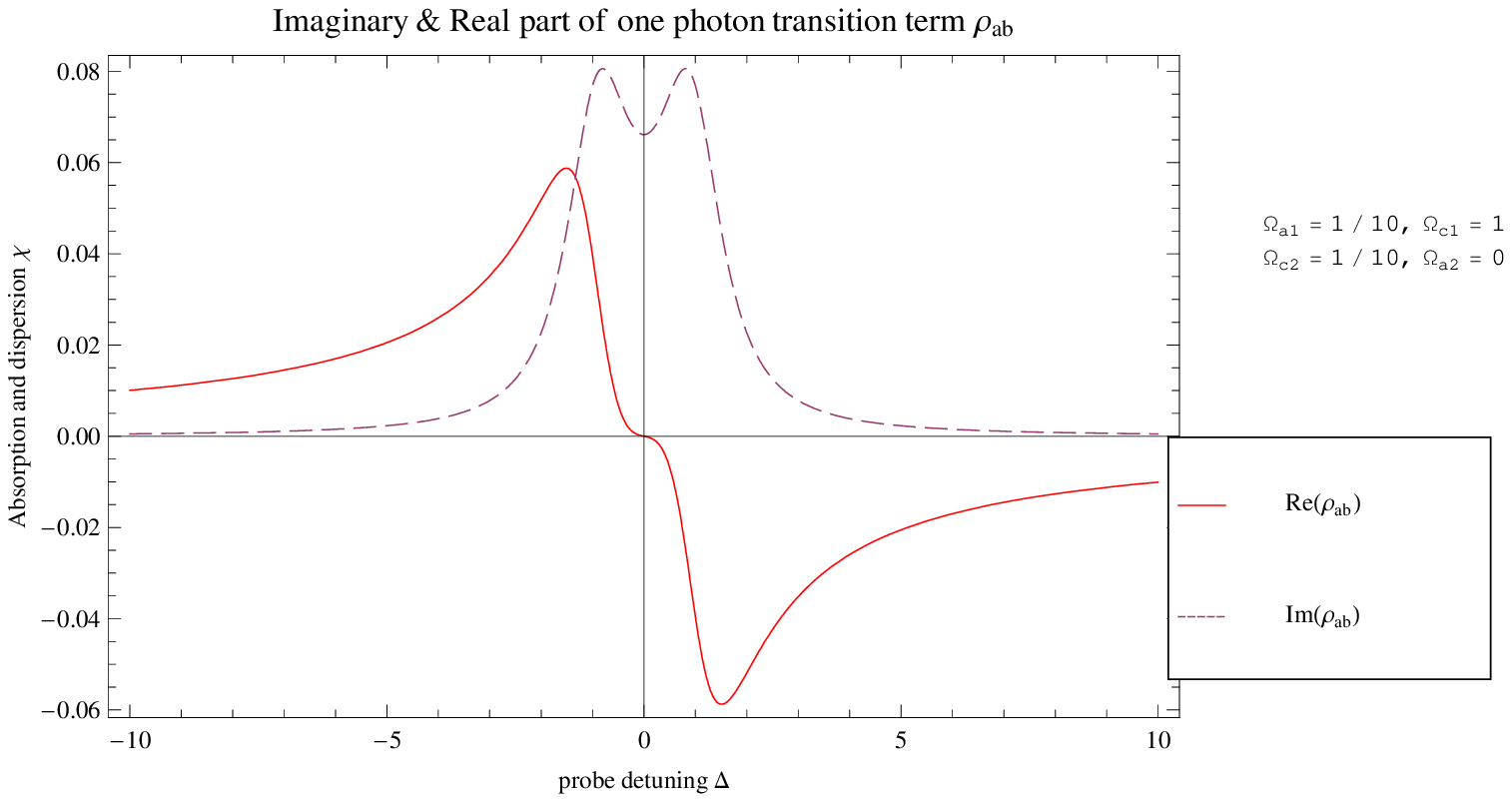}\\ \hline
 \end{tabular}
\caption{\label{fig:probe2} For the 2nd symmetry broken case, the
probe term $\rho_{cd}$ (figure a and a'), trig transition term
$\rho_{ab}$ (figure b \& b') vary with field detuning $\Delta$,
respectively. The parameters for situation in the left column
(figure a, b) are: $\Omega_{a1}=1/10, \Omega_{a2}=0, \Omega_{c1}=5,
\Omega_{c2}=1/10$, all the decay rates take $\gamma_i=1(i=1,2,3,4)$,
all detunings take zero value except the probing one. We decrease
the strength of the couple field to $\Omega_{c1}=1$ for situation on
the right column (figure a', b'), other parameters are remained
unchanged as the left column.}
 \end{figure}

\begin{figure}
 \begin{tabular}{|c|c|}\hline
 c\includegraphics[width=0.45\columnwidth]{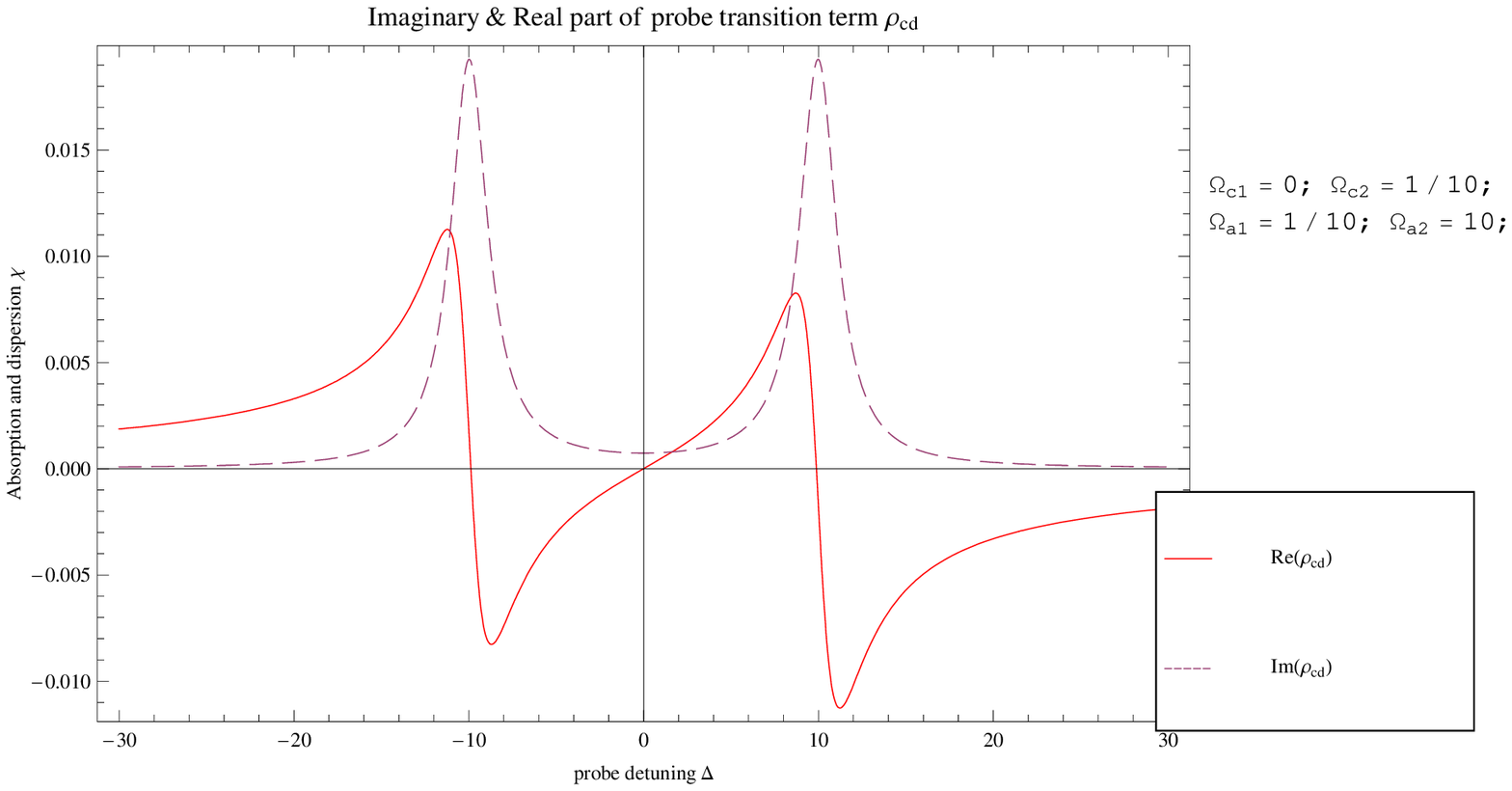}&
 c'\includegraphics[width=0.45\columnwidth]{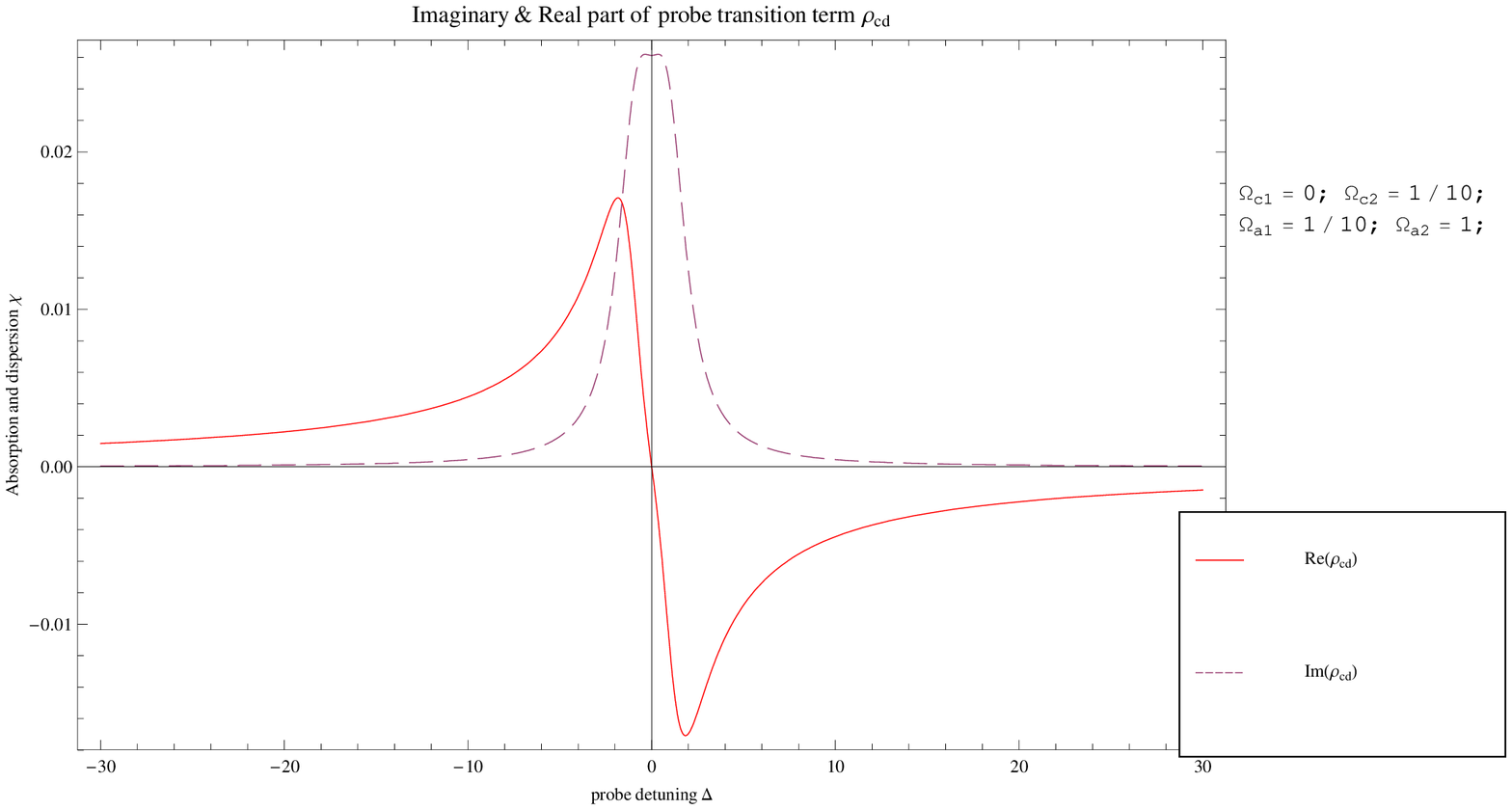}\\ \hline
 d\includegraphics[width=0.45\columnwidth]{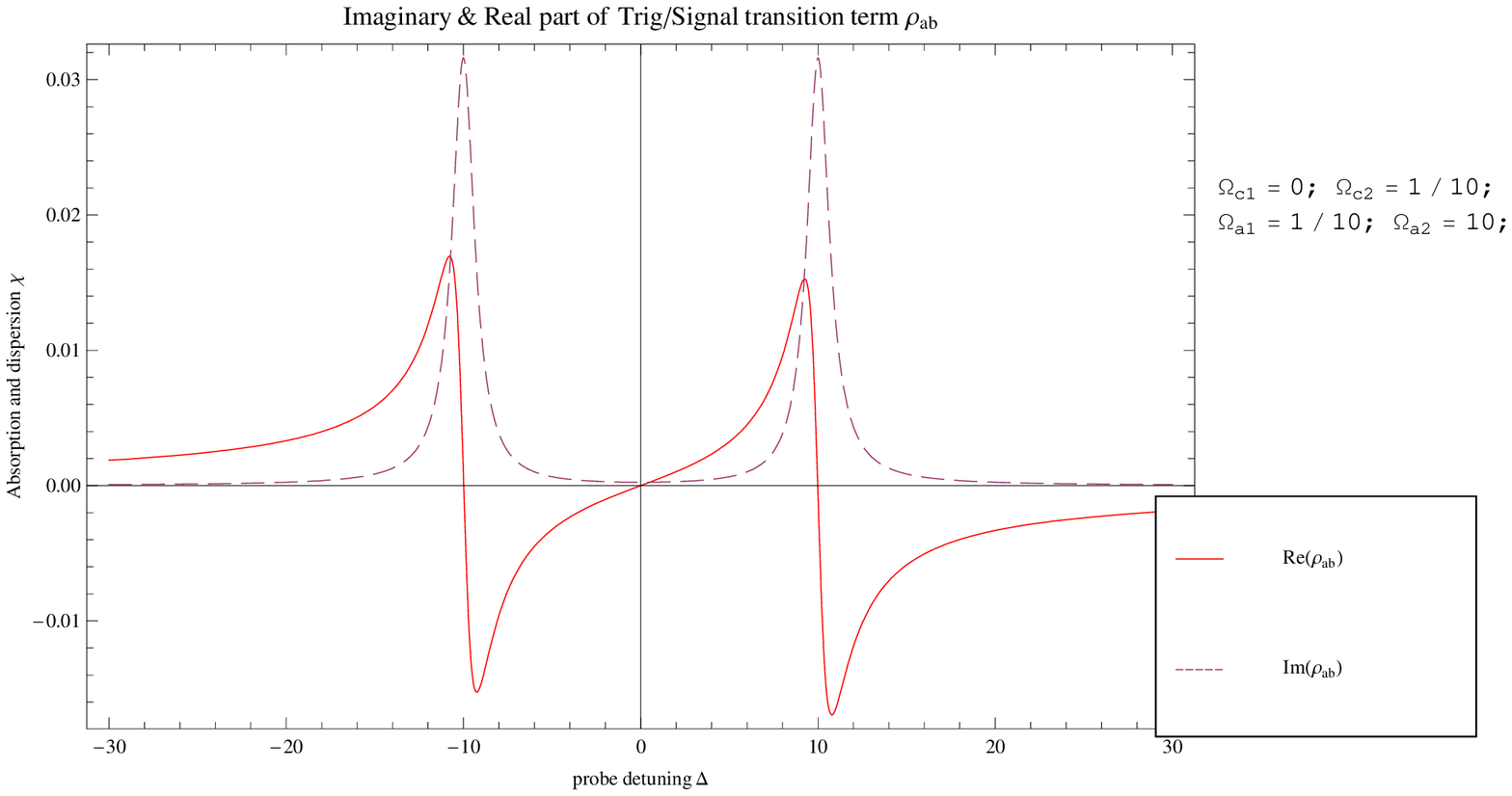}&
 d'\includegraphics[width=0.45\columnwidth]{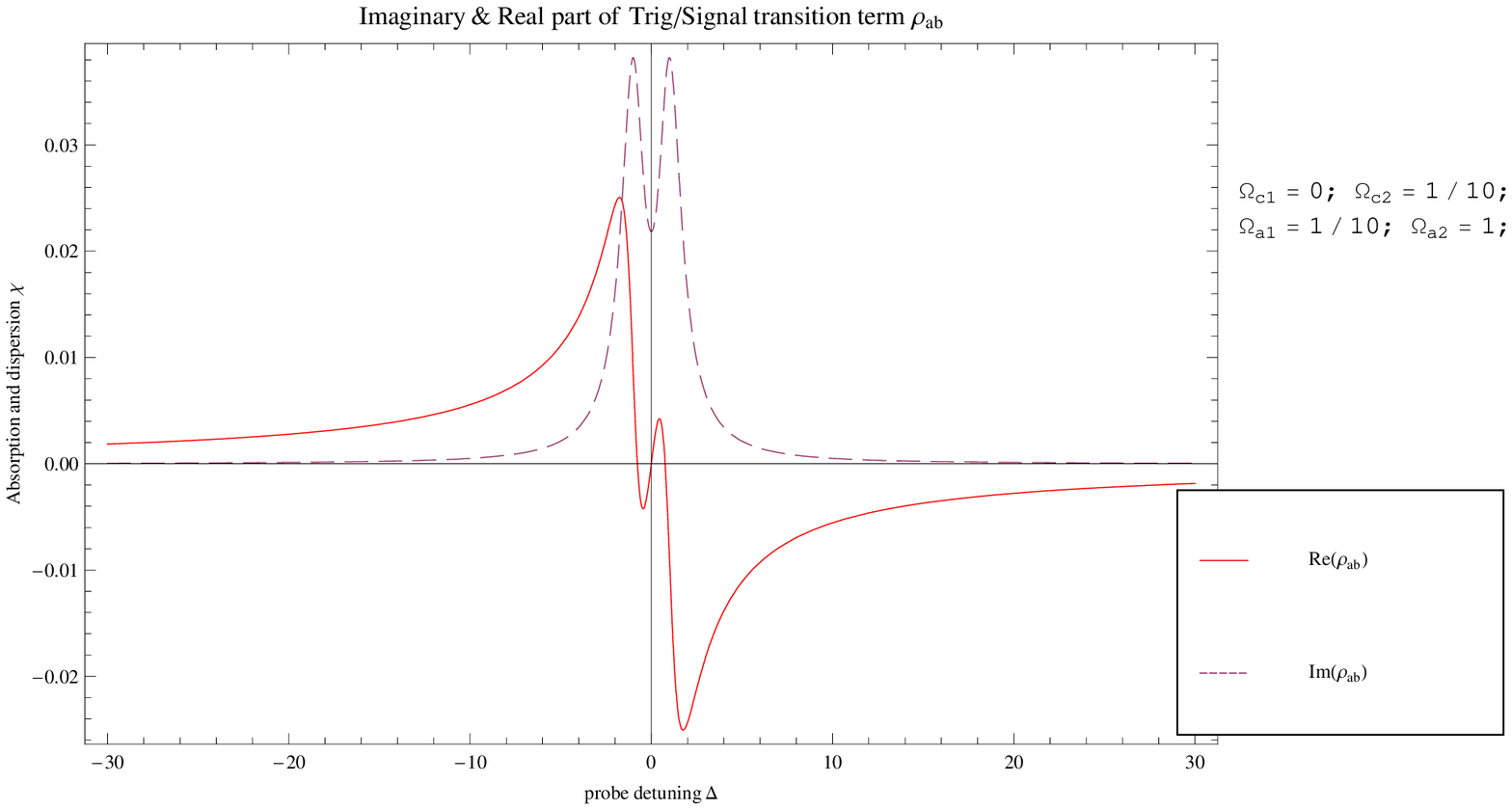}\\ \hline
 \end{tabular}
\caption{\label{fig:probe3} For the 3rd symmetry broken case, the
probe term $\rho_{cd}$ (figure c \& c'), the trig terms $\rho_{ab}$
(figure d \& d') vary with field detuning $\Delta$, respectively.
The parameters for left column (figure c, d ) are:
$\Omega_{a1}=1/10, \Omega_{a2}=10, \Omega_{c1}=0, \Omega_{c2}=1/10$,
all the decay rates take $\gamma_i=1$, all detunings take zero value
except the probing one. And the strength of the couple transition
decreases to: $\Omega_{a2}=1$ for situation on the right column
(figure c', d'), other parameters are the same as those in the left
column.}
 \end{figure}

 While for the second and the third kind of symmetry broken
situation, where dark states resonance will also appear, the same
analysis method can be applied, that is, by investigating the
level-population, the absorption and dispersion of the probing
transition, and the two-photon or three-photon transition terms, the
quantum interference and quantum coherence features of the
four-level atomic system interacting with multi-optical fields will
be revealed clearly. Here the absorption and dispersion behaviors of
the probe and trig fields transition terms are shown in
Fig.\ref{fig:probe2} and Fig.\ref{fig:probe3}, respectively. Let's
discuss the second symmetry broken case a little further, where the
diamond atomic system is driven by three coherent optical fields
$E_1, E_2$ and $E_3$. For convenience, we denote the field $E_1$ as
the couple field, $E_2$ the probe field and $E_3$ the trig field. In
the Fig.\ref{fig:probe2}, the absorption and dispersion characters
of the three fields are drawn for two kinds of different coupling
strength. On the left row, the coupling strength ($\Omega_{c1}=5$)
is much stronger than that of the probe one ($\Omega_{c2}=1/10$),
and interestingly the EIT effects for the probe field and the trig
field appear at the same time(figure a, b); but if the strength of
the couple field decreases (on the right row of
Fig.\ref{fig:probe2}), the depth of the transparency windows for the
probe and the trig fields will become smaller, and the transparency
characters of both fields will vanish when the strength of the
couple field takes $\Omega_{c1}=1$ (figure a' and b'), thus the
couple field can \textquotedblleft{control}\textquotedblright the
behaviors of the probe and trig fields. And for the third kind of
symmetry broken case, similar characters are shown in
Fig.\ref{fig:probe3}, where the double-EIT effects for the probe and
trig fields are controlled by the strength of the couple field.
These interesting controlled double-EIT effects root in the
interacting dark states resonances, and will promise interesting
applications on quantum optics and quantum information science.

\section{\label{sec:Diss} Discussion and Conclusion}

In the above, we have studied the quantum interference effects of
the $\diamondsuit$ four-level atomic system, and revealed various of
dark states and the resulted electromagnetically induced
transparency effects. As is well known that, the EIT system is
tightly related to the enhancement of nonlinearity, and multi-level
EIT system is efficient to gain the large nonlinearity and large
cross-phase-modulation(XPM), such as the
\textquotedblleft{N}\textquotedblright type four-level
system\cite{OL1936, JMO1559}, the
\textquotedblleft{M}\textquotedblright type four-level
system\cite{PRL197902}, the tripod type\cite{PRA032317} and the
inverted-Y type\cite{PRA062319} four-level system. The XPM effect is
very important in implementing two-qubits all-optical quantum phase
gate(QPG). When the two qubits are defined as two polarized optical
pulses: a probe light pulse and a trigger one, the phase gate is
implemented by the cross-phase-modulation between the two pulses. If
these two optical pulses are able to go through an enhancement of
large nonlinearity simultaneously, then the XPM process, the quantum
phase gate operation will be achieved. Thus the double-EIT effects
of the probe and trigger optical fields is necessary to realize the
QPG\cite{OL1936, JMO1559, PRL197902, PRA032317, PRA062319}. The
$\diamondsuit$ system we are discussing here is a new type of
multi-level EIT system differs from the systems studied in the above
Ref.\cite{OL1936, JMO1559, PRL197902, PRA032317, PRA062319}, though
the detail of implementing the QPG operation by this $\diamondsuit$
system remains to further study, it is confirmable to say that the
diamond system will be useful to quantum information science.
Comparing with all the other atomic structures in the above
references, the $\diamondsuit$ system is a kind of more subtle
multi-level system, and it has several advantages. Firstly, the
diamond system is a very general system, the atomic levels and the
four coherent optical driving fields are nearly free of restriction,
so it would be easier to realize experimentally. Secondly, there are
many ways of arranging the roles of the probe field and the trigger
field, that is, all four fields are able to assign to be the probe
and and the trigger field. Finally, we are aware that an experiment
of coherently controlling the $\diamondsuit$ system in cold
$^{87}{Rb}$ has been carried out recently\cite{PRL203001}, which
demonstrates the feasibility of implementing the quantum phase gate
operation. This part of contents are worth to study in the future.

In summary, we have explored the rich quantum interference and
coherence effects of the $\diamondsuit$ system driven by four
coherent optical fields, and reveal the origin of the attractive
effects such as electromagnetically induced transparency. By
comparing to other type of atomic system, we demonstrate that the
diamond system is much more accessible experimentally, which
promises charming applications in quantum optics, nonlinear optics
and quantum information science.

\begin{acknowledgments} This work was supported by National Funds of
Natural Science of China (Grant No. 10504042). \end{acknowledgments}


\end{document}